\documentclass[doublecol,figures]{epl2} 

\title{Electron energy loss spectroscopy determination of Ti oxidation state at the (001) LaAlO$_3$/SrTiO$_3$ interface as a function of LaAlO$_3$ growth conditions}
\shorttitle{EELS determination of Ti oxidation state at the LaAlO$_3$ / SrTiO$_3$ interface} 

\author{J.-L. Maurice\inst{1} \and G. Herranz\inst{1} \and C. Colliex\inst{2} \and I. Devos\inst{3} \and C. Carr\'et\'ero\inst{1} \and A. Barth\'el\'emy\inst{1} \and K. Bouzehouane\inst{1} \and S. Fusil\inst{1} \and D. Imhoff\inst{2} \and \'E. Jacquet\inst{1} \and F. Jomard\inst{4} \and D. Ballutaud\inst{4} \and M. Basletic\inst{1}}
\shortauthor{J.-L. Maurice \etal}

\institute{                    
  \inst{1} Unit\'e mixte de physique CNRS/Thales and Universit\'e Paris-Sud-XI, Route d\'epartementale 128, 91767 Palaiseau cedex, France\\
  \inst{2} Laboratoire de physique des solides CNRS UMR8502, Universit\'e Paris-Sud-11, 91405  Orsay cedex, France\\  
  \inst{3} Institut d'\'electronique, de micro\'electronique et de nanotechnologie, Unit\'e Mixte de Recherche CNRS 8520, avenue Poincar\'e BP 69, 59652 Villeneuve d'Ascq cedex, France\\
  \inst{4} GEMaC-UMR 8635, CNRS, 1 place Aristide Briand, 92195 Meudon cedex, France\\
}

\abstract{
At the (001) interface between the two band-insulators LaAlO$_3$ and SrTiO$_3$, a high-mobility electron gas may appear, which has been the object of numerous works over the last four years. Its origin is a subject of debate between the interface polarity and unintended doping. Here we use electron energy loss 'spectrum images', recorded in cross-section in a scanning transmission electron microscope, to analyse the Ti$^{3+}$ ratio, characteristic of extra electrons. We find an interface concentration of Ti$^{3+}$ that depends on growth conditions.}

\begin{document}

\maketitle

\section{Introduction}
In a pioneering work, Ohtomo and Hwang\cite{Ohtomo2004} have discovered that growing LaAlO$_3$ (LAO) by pulsed laser deposition (PLD) onto a TiO$_2$-terminated (001) surface of SrTiO$_3$ (STO), produces a metallic system, while both materials are insulators with respective band gaps of 5.6 eV and 3.2 eV\cite{Robertson2004}. This observation has been the signal for many studies of this system: among remarkable results are the superconductivity observed below 200 mK\cite{Reyren2007}, the magnetism around 300 mK\cite{Brinkman2007}, and the field effect on conductivity\cite{Thiel2006}. 

In the ionic limit, the perfect interface carries a global positive charge, because the last TiO$_2$ plane of STO is neutral and faces a first plane of LAO that has composition LaO and carries an uncompensated charge +1/2 per surface unit cell. Such a configuration would lead to a so-called 'polar catastrophe'\cite{Noguera2000} if it were not balanced by electron or ion rearrangements. In the present case, the capacity of Ti ions to bear a mixed valence may provide the necessary screening, some Ti$^{4+}$ (valence in bulk STO) becoming Ti$^{3+}$. In this most simple scheme -- essentially confirmed by calculations \cite{Pentcheva2006,Park2006,Maurice2007} --  $1/2$ extra electron per surface unit cell is necessary to maintain electrical neutrality\cite{Ohtomo2004}. The spreading of the electron cloud associated with such a delta-doping has been evaluated by solving Poisson's equation: at 300 K, electron concentration should be $\sim$1.3 $\times$ 10$^{21}$ cm$^{-3}$ ($\sim$8 $\%$ of the Ti concentration) at the interface, and it should fall to $\sim$8 $\times$ 10$^{20}$ cm$^{-3}$ (5$\%$, the present detection limit, see below) at only 1 nm into the bulk\cite{Siemons2007}.

Several authors have associated the n-type conductivity they have measured with such screening electrons\cite{Ohtomo2004,Thiel2006,Huijben2006}. However, growth by PLD is a complex process, which will hardly create 'perfect' materials\cite{Ohnishi2005,Willmott2007}. Thus this scheme of interpretation, based on the interface being atomically sharp and chemically stoichiometric, is the object of intense discussion. In our laboratory, we have inferred from Shubnikov-De-Haas oscillations\cite{Herranz2007} and cross-section conducting-tip atomic-force microscopy (CT-AFM)\cite{Basletic2008}, that, in the samples grown at low-pressure, which exhibited the highest mobilities, the electron gas was 3-dimensional and extended hundreds of $\mu$m into the bulk of the substrate. Moreover, Siemons \etal\cite{Siemons2007}, by making measurements before and after oxygen annealing, and Kalabukhov \etal\cite{Kalabukhov2007}, by measuring cathodo- and photoluminescence, concluded that the doping was due to oxygen vacancies that had diffused during the PLD process. In reality, pulsed laser deposition proceeds with species (atoms, ions and clusters) that have a kinetic energy upon landing which depends, among other growth parameters, on the pressure in the chamber: in our case, typically several hundreds of eV in the 10$^{-2}$ - 10$^{-4}$ Pa pressure range, and decreasing at higher pressure\cite{Khodan2007}. The point defect content of the growing layer thus directly depends on the growth oxygen pressure, not merely because vacuum creates vacancies, but mainly because landing species have enough energy to create irradiation damage. This in turn, makes the growing layer a pressure-dependent reservoir of vacancies for exodiffusion from the substrate. This phenomenon of substrate reduction is however well known in molecular beam epitaxy, where it is utilised for the oxidation of epitaxial layers\cite{Uedono2002}.

If bulk substrate properties thus depend on the growth pressure of the epitaxial layer, the question remains about the interface role on conductivity. With our CT-AFM experiments\cite{Basletic2008}, we have shown that the interface region did exhibit a specific, enhanced, conductivity, but -- and this is a key point -- with a low electron mobility. Correlatively, we also found that this conductivity depended on growth conditions\cite{Basletic2008}. Low-mobility and dependence on growth conditions would be both the logical result of an extrinsic source of dopant. It is thus mandatory to characterise the doping level at the interface itself, as a function of growth conditions. This need is further highlighted by a recent result based on X-ray diffraction\cite{Willmott2007} that shows the existence of a several unit-cell thick layer of metallic La$_{1-x}$Sr$_{x}$TiO$_3$ at the interface.

Our high-resolution transmission electron microscopy (HRTEM) observations in cross section have shown the existence of a distortion at the interface\cite{Maurice2006b}, which could be equally consistent with the delta-doping \cite{Park2006,Maurice2007} or the La$_{1-x}$Sr$_{x}$TiO$_3$ hypothesis\cite{Willmott2007}. But they also exhibited noisy contrasts \cite{Maurice2007} that would finally back up the latter. Measurements of the  Ti$^{3+}$ ratio were carried out by electron energy loss spectroscopy (EELS) in the scanning transmission electron microscope (STEM), by David Muller's group\cite{Nakagawa2006,Reyren2007} and ours\cite{Maurice2007}. The LAO/STO interfaces analysed were in all these cases prepared at relatively low pressures; Ti$^{3+}$ was detected at the interface, over a width reaching 5 nm in ref.~\cite{Nakagawa2006} and closer to the nm in refs.~\cite{Reyren2007,Maurice2007}, thus depending, very likely, on the details of sample growth and preparation.

Here, we bring new STEM-EELS data, obtained this time from a sample grown at 40 Pa (0.4 mbar, 'high pressure'), which was macroscopically insulating, and from a new reference grown at 10$^{-4}$ Pa (10$^{-6}$ mbar, 'low pressure'), macroscopically conductive. Our results evidence, for the first time to our knowledge, the role of growth conditions on the microscopic interface atomic structure: oxygen pressure in the growth chamber tends to decrease the apparent interface doping. The latter is therefore not only due the intrinsic polarity, but also to extrinsic point defects. We have additionally performed a combined HRTEM, electron-diffraction, and atomic force microscopy study of the two samples, which shows a fundamental change in relaxation of the epitaxial strain as the growth pressure increases. At low pressure, relaxation takes place without dislocations in a way preserving a 2-dimensionnal layer by layer growth, most probably with the aid of point defects. At high pressure, relaxation occurs through a 3-dimensionnal growth mode.

\section{Methods}
LAO thin films were grown by pulsed laser deposition (PLD) using a frequency-tripled ($\lambda$ =355 nm) Nd: yttrium aluminum garnet (YAG) laser on TiO$_2$-terminated STO substrates [Surfacenet Gmbh] with growth oxygen pressures PO$_2$ = 10$^{-4}$ Pa and PO$_2$ = 40 Pa, for the low- and high-pressure specimens, respectively, at a deposition temperature of T = 750 $^{\circ}$C. The laser pulse rate was fixed at 2.5 Hz, whereas the energy density was 2.8 Jcm$^{-2}$. The target-substrate distance was 55 mm, resulting in a growth rate of about 1 \AA s$^{-1}$. Once the film deposition was finished, the samples were cooled down to room temperature at the growth oxygen pressure (10$^{-4}$ and 40 Pa, respectively). The transport properties of the low-pressure sample are among the best published to date (mobility $>$ 10$^4$ cm$^2$V$^{-1}$s$^{-1}$ at 4K, see details in ref.~\cite{Herranz2007}). In that specimen, the LAO thickness was 20 nm. The other sample was insulating, the LAO in this case was 5-nm thick. This choice was made so that point-defect diffusion processes would dominate in the low-pressure sample\cite{Herranz2007, Basletic2008} and be negligible in the high-pressure case. 

The LAO/STO cross-sectional samples for TEM and STEM-EELS were prepared using standard tripod polishing, without water for the last $\sim$5 $\mu$m on each face. It was completed by a prolonged, grazing incidence, ion milling with 2-keV Ar ions, from the backside (STO) only. This procedure resulted, as shown by the profiles below (see fig.~\ref{fig.5}), in thin foils with fairly parallel faces. This point is very important as the surfaces of foils are strong sources of Ti$^{3+}$. We otherwise minimised the influence of surfaces by selecting relatively thick regions ($\sim$30-40 nm). We checked by HRTEM the defects created by STEM-EELS recording, and kept only those data sets where they remained invisible. The STEM-EELS measurements were performed with a VG HB501 microscope operated at 100 kV, TEM and HRTEM observations were carried out on a Topcon 002B, working at 200 kV.

We analysed the spatial variations of electron energy loss spectra using spectrum images (3-dimensional data boxes that consist in $n$ $\times$ $m$ arrays of $p$-channel spectra\cite{Jeanguillaume1989}) acquired by scanning a focused probe step by step and recording a spectrum at each step. We typically recorded images consisting in 2 to 8 64-pixel lines perpendicular to the interface, with a pixel size of 0.33 nm to 0.74 nm and a channel size of 0.2 eV. The dwell time varied between 0.2 and 1 s. In order to increase signal/noise ratio, the profiles finally used, such as those presented below, were obtained by summing, parallel to the interface, the 2 to 8 lines. Before this, the drift, measured as the shift of the interface from one line to the next, was corrected on pixel position and size. The effective size of the interaction volume was calculated by deconvoluting the measured Ti concentration profiles, assuming a step function for the actual profiles. The intensity distribution of the probe thus defined was quasi-gaussian with a width at half maximum of $\sim$2nm in the experiments presented below. The energy resolution of the spectrometer was better than 0.6 eV (width at half maximum of zero-loss peak). 

\begin{figure}
\onefigure[width=6cm]{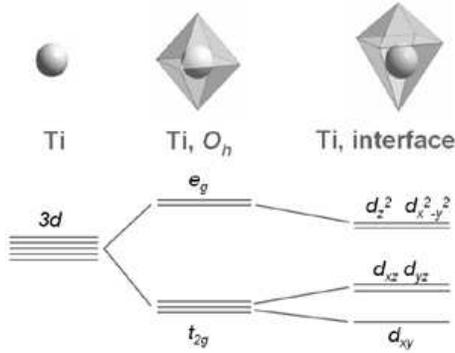}
\caption{Evolution of the $3d$ electronic structure of Ti as a function of valence and surroundings. Degeneracy in the case of Ti$^{4+}$ in bulk STO is partially lifted by the octahedral crystal field (middle), separating $3d$ into $t_{2g}$ and $e_g$ levels. The two corresponding peaks are well visible in EELS Ti-$L_2$ and Ti-$L_3$ edges (see \textit{A, B} in fig.~\ref{fig.4})\cite{Maurice2006a,Abbate1991}. In the case of Ti at the interface (right), both valence and site-symmetry are different, which organises the $3d$ levels in quite a different way\cite{Maurice2007}. The levels presented correspond to molecular calculations, an additionnal spreading will occur in the actual crystal. This will transform the two peaks into a broad one in the EELS edges.}
\label{fig.1}
\end{figure}

The evaluation of the Ti$^{3+}$ ratio was carried out using the high sensitivity on this ratio of the fine structure of Ti-$L_{2,3}$ edge\cite{Abbate1991,Ohtomo2002,Maurice2006a}. Indeed, this edge corresponds to transitions from $2p$ discrete levels to the $3d$ band, which is empty for Ti$^{4+}$ in bulk STO, and contains one half an electron per site in the simplest model of Ti$^{3+/4+}$ at the interface. Fig.~\ref{fig.1} compares the Ti-$3d$ molecular electronic structure in the case of the octahedral crystal field of STO and that of the distorted site\cite{Maurice2007} at the abrupt LAO-STO interface. The move of energy levels, when going from bulk STO to the interface, will appear in spectra as a fine structure smoothing. 

More specifically, both the Ti-$L_2,$ and Ti-$L_3$ edges in STO are made of two well defined peaks corresponding to the $e_g$ and $t_{2g}$ levels separated by the octahedral crystal field (see fig.~\ref{fig.4} and ref.~\cite{Maurice2006a}): a shift a valence towards 3+ will show as a progressive fading of these peaks and an  increase of signal in the valley in between (see arrows in fig.~\ref{fig.4}). It is not possible, of course, to separate geometry from valence contributions in such an evolution. However, as the two are intrinsically related through Jahn-Teller type effects, in the analysis presented below, we have considered that all changes of line shape were associated with a valence change. 

We further assumed that the global Ti-$L_{2,3}$ cross section did not depend on the oxidation state of Ti, so that all experimental spectra be linear combinations of pure Ti$^{4+}$ and Ti$^{3+}$ spectra. We performed least squares fits of the experimental Ti-$L_{2,3}$ edges with reference Ti$^{4+}$ and Ti$^{3+}$ spectra. The fits were performed after background subtraction, and after normalising the integral of the remaining signal. The reference Ti$^{4+}$ spectrum was picked from the substrate contribution in the same profile where the analysed spectrum came from and the Ti$^{3+}$ spectrum was taken in ref.\cite{Ohtomo2002}.  

In such measurements, sources of error are numerous\cite{Colliex1989}. In order to increase the signal/noise ratio, we also applied the least-squares analyses to Ti-$L_{2,3}$ edges obtained after summing as many comparable interface spectra as possible. The summed edges totalised more than 2.5 $\times$ 10$^6$ counts (background subtracted), in both the cases of low and high growth pressure (see fig.~\ref{fig.4}). Even in such conditions, the error in our measurements, taken as the standard deviation between data points from the experimental and fitting spectra, remained of the order of 3 $\%$, which made in turn our detection limit (3 $\%$: $\sim$5 $\times$ 10$^{20}$ cm$^{-3}$).

\begin{figure}
\onefigure[width=7.5cm]{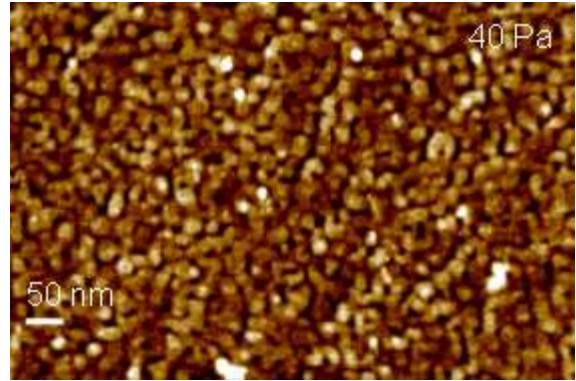}
\caption{Atomic force micrograph of the high-pressure sample exhibiting 3-D features. In the area shown, the largest peak-to-valley depth reaches 6 nm, i.e. the film thickness.}
\label{fig.2}
\end{figure}

\section{Point defects in LAO}
We show in the present section that the two samples have undergone different relaxation processes of the epitaxial strain, which we associate with different point-defect contents. We focus in the following on the high pressure sample, where growth appears to be quite different from that usually observed in the LAO/STO system. 

\begin{figure}
\onefigure[width=8.5cm]{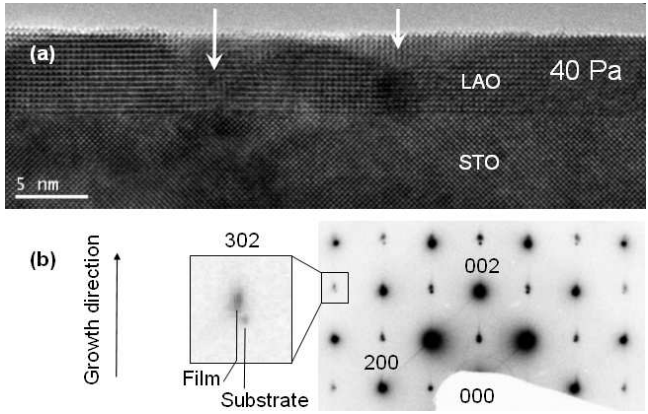}
\caption{HRTEM (a) and corresponding selected area electron diffraction pattern (b) of the high pressure sample. As can be seen with the lateral shift of 302 spot, the in-plane parameter is different in film and substrate, which indicates plastic relaxation, a process that does not occur in the low-pressure sample (not shown, see e.g. ref.~\cite{Maurice2006b}). The arrows in (a) indicate valleys between relaxed 3D regions.}
\label{fig.3}
\end{figure}

\begin{figure}
\onefigure[width=7.5cm]{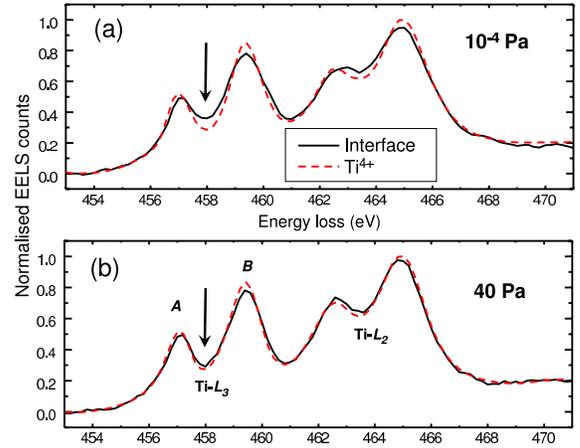}
\caption{Electron energy loss Ti-L$_{2,3}$ edge recorded at the interface in both the low-(a) and high-(b) pressure samples, compared in each case with a bulk spectrum recorded nearby, giving the reference Ti$^{4+}$. These normalised spectra were obtained after summing several recordings, see text. Note the two peaks \textit{A} and \textit{B} corresponding to $t_{2g}$ and $e_g$ levels, and the signal in the valley (arrows), significantly more important in the low-pressure case.}
\label{fig.4}
\end{figure}

Figure~\ref{fig.2} shows an atomic force micrograph of this sample where the surface appears significantly rough, contrary to what is generally observed with low pressure growth. We associate this roughness with 3-dimensionnal growth. The strucural analysis of this high-pressure sample by TEM in $<$100$>$ cross section (fig.~\ref{fig.3}) indicates that some plastic relaxation has occurred. Selected area electron diffraction (fig.~\ref{fig.3}b) shows that the distorted pseudo-cubic parameters $a$ (in-the-plane) and $c$ (out-of-plane) are both modified. Taking the internal reference given by STO ($a_{STO}$ = 0.3905 nm), the lattice pseudo-cubic parameters of this sample come out at $a_{LAO}$ = 0.3873 nm and $c_{LAO}$ = 0.3748 nm (+/- 0.001 nm). Correlatively with the in-plane mismatch, we found some misfit dislocations, but not enough to allow for all the difference measured. Plastic relaxation thus appears to be associated with the 3-dimensional growth mode. When compared to the equilibrium pseudo-cubic lattice parameter of LAO (0.3792 nm), the parameters found indicate a volume extension, which may be attributed to an elastic distortion with a Poisson ratio of about 0.22. 

In similar TEM experiments carried out on the low-pressure sample (not shown), the LAO lattice parameters came out at $a_{LAO}$ = 0.3905 and  and $c_{LAO}$ = 0.3818. The low-pressure film thus appeared to be fully strained to the substrate, with a relatively small elastic relaxation out-of-plane as has been noticed previously\cite{Reyren2007,Maurice2006b}. These values indicate a unit cell volume much larger than in the high pressure case, which is consistent with the presence of a much larger amount of point defects. And indeed, in the observation of the $<$110$>$ cross section of a sample grown in identical conditions (not shown), we have evidenced the existence of a superstructure which could not be simulated using the exact equilibrium rhombohedral LaAlO$_3$ unit cell, and had to be due to ordering of large amounts of point defects.

Therefore, the two samples would strongly differ in point-defect content, consistently with their growth conditions: the low-pressure sample having been hit by fast species during growth would contain a much higher level of point defects than the high-pressure one. 

\section{EELS results} 
\begin{figure}
\onefigure[width=7cm]{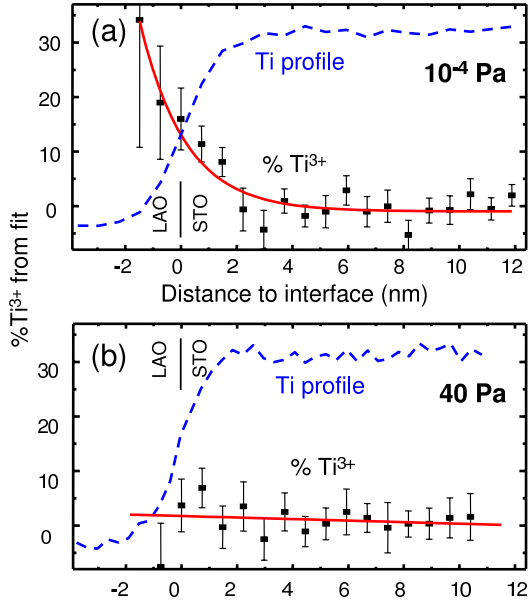}
\caption{(colour online) Ti$^{3+}$ profiles in the low-(a) and high-(b) pressure samples. Concentrations obtained after fitting all spectra in each profile with Ti$^{3+}$ and Ti$^{4+}$ references as described in text. Superimposed are the corresponding normalised Ti profiles in each case. The red curves are guides to the eye.}
\label{fig.5}
\end{figure}

Fig.~\ref{fig.4} presents raw experimental spectra, of which the treatment has been limited to background subtraction and averaging over several spectrum-images. In fig.~\ref{fig.5}, the data have been deduced from the spectra by the fitting procedure described in sec. 'Methods'. The experimental conditions were quite similar (probe effective size and current, spectrometer setting), so that the two samples in each figure can be directly compared. 

In fig.~\ref{fig.4}, the mean experimental interface spectra are superimposed on bulk spectra, giving the reference Ti$^{4+}$. The effect of the interface on the Ti-L$_{2,3}$ edge is clearly visible in the low pressure sample, while it is significantly smaller in the high-pressure case. The fit with reference Ti$^{3+}$ and Ti$^{4+}$ spectra as described in sec. 'Methods' gives 20$\%$ Ti$^{3+}$ in the former case and only 10 $\%$ Ti$^{3+}$ in the latter. 

Fig.~\ref{fig.5} shows the spatial evolution of Ti$^{3+}$ ratio, deduced from such fits, in two profiles taken in the low- (fig.~\ref{fig.5}a) and high- (fig.~\ref{fig.5}b) pressure samples. Error bars were set to twice the standard deviation between experimental and fitting spectra. Variations from one profile to the next were negligible in the low-pressure sample, so that the profile shown in fig.~\ref{fig.5}a is quite representative of this sample. In contrast, such variations were quite significant in the high-pressure sample, probably due to the 3-dimensional relaxation of strain in LAO and its localised effects in the substrate. For the profile presented in fig.~\ref{fig.5}b we have chosen a region where the interface Ti$^{3+}$ level was the lowest. It appears on the figure to be of the order of uncertainty, which is about 5 $\%$ in this case. 

In order to help reading fig.~\ref{fig.5}, let us recall that the probe has a gaussian shape with a width at half maximum of 2 nm. Thus, it starts to 'count' Ti and Ti$^{3+}$ when it is still centered in pure LAO. This causes the residual Ti profile in LAO, but also, as the proportion of Ti$^{3+}$ is the largest at that time, the shift towards LAO of the maximum of Ti$^{3+}$ ratio in fig.~\ref{fig.5}a. Because of these probe convolution effects, it is necessary to compare the present profiles with models to effectively localise the Ti$^{3+}$ ions, which we do in the next section. 

In turn, as measurement noise is the largest where the Ti signal is the smallest, the actual uncertainty there (see first data point in fig.~\ref{fig.5}b) is probably larger than our estimate based on fit quality.

\begin{figure}
\onefigure[width=7cm]{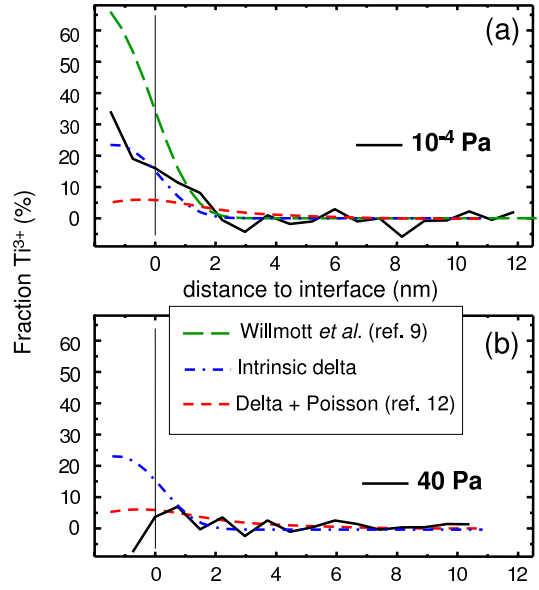}
\caption{The Ti$^{3+}$ profiles of fig.~\ref{fig.5} in the low-(a) and high-(b) pressure samples compared with the EELS response applied to published Ti$^{3+}$ distributions. 'Intrinsic delta': localised electrons that would neutralise the intrinsic polarity in the form of a delta distribution at the interface; 'Delta + Poisson': same surface density of electrons, but spread into the bulk following the solution of Poisson equation\cite{Siemons2007}; Willmott\etal : distribution of Ti$^{3+}$ given in fig. 3a of ref.~\cite{Willmott2007}.}
\label{fig.6}
\end{figure}

\section{Discussion}
We compare in this section our experimental profiles with different models of electron distributions, convoluted with our gaussian experimental response (fig.~\ref{fig.6}). Let us first note that, once the Ti$^{3+}$ profiles from the literature are convoluted, they only slightly emerge from the present detection limit (for the present profiles, the sensitivity is worth $\sim$5 $\%$ or $\sim$8 $\times$ 10$^{20}$ cm$^{-3}$, of course not as fine as that mentioned above for summed spectra). Thus it is not only justified, but even mandatory, to include our high pressure profile, even though it is almost flat, in such comparisons.

Quite surprisingly, in both our cases of low- (fig.~\ref{fig.6}a) and high- (fig.~\ref{fig.6}b) pressures, the best fit is obtained with a model of intrinsic polarity. However, the low-pressure -- conducting -- case appears to fit with a model of localised interface electrons ('Intrinsic delta' in fig.~\ref{fig.6}a), while the high pressure -- insulating -- case fits on the contrary with a model of mobile electrons ('Delta + Poisson' in fig.~\ref{fig.6}b). As this is precisely the opposite of what is expected, we have to attribute an extrinsic contribution to at least one of the two cases.

Given that electrons could hardly stay localised in the conducting low pressure specimen, and given that we have made it so as to increase extrinsic effects, we may appoint it to represent the extrinsic case. Thus, we would rather take the present sharp profile of fig.~\ref{fig.6}a as the tip of a broad distribution of mobile carriers such as that we present in ref.~\cite{Basletic2008}. If we recall that standard carrier densities are orders of magnitude lower\cite{Herranz2007, Basletic2008} than the present detection limit ($\sim$8 $\times$ 10$^{20}$ cm$^{-3}$), this hypothesis appears quite consistent. Thus this low-pressure profile would be associated with donor-like point defects such as oxygen vacancies or La substituted for Sr. But fast kinetics of oxygen exodiffusion\cite{Uedono2002} appear uncompatible with the profile sharpness, and allow one to eliminate oxygen vacancies. Consequently, we checked cation interdiffusion by secondary ion mass spectrometry on much thicker samples. And we found indeed an enhanced La-Sr exchange, compared to Al-Ti, that moreover depended on growth pressure\cite{Carretero2008}. The most likely dopant would thus be Lanthanum rather than the oxygen vacancy. It can be noticed, by the way, that Sr$_{1-x}$La$_{x}$TiO$_{3}$ chemical solutions are stable over a large concentration range\cite{Eror1981,Tokura1993}. Thus, we would finally have an extrinsic doping effect, of the type evoked by Willmott\etal\cite{Willmott2007} (fig.~\ref{fig.6}a), though in lower amounts.

In the high-pressure case (fig.~\ref{fig.6}b), the Ti$^{3+}$ profile should indeed be closer to a model of intrinsic polarity. But it remains quite surprising that it be the mobile-carrier version of the model, as this sample is macroscopically insulating. An explanation, perhaps difficult to check, would be that the sample is indeed conductive in between the areas where epitaxial stress is concentrated (arrows in fig.~\ref{fig.2}). The stressed zones would, in turn, act as traps and quench conductivity.

\section{Conclusion}
We have measured the concentration profiles of Ti$^{3+}$ ions in the STO substrate in the vicinity of the TiO$_2$-terminated (001) interface with LAO using STEM-EELS in cross section. We have studied two growth conditions: 200 s and 10$^{-4}$ Pa on the one hand, and 50 s and 40 Pa on the other hand, so that point-defect effects be maximum in the former case, and minimum in the latter. The concentration of Ti$^{3+}$ ions appears to depend significantly on growth conditions: in the low-pressure, thicker sample, Ti$^{3+}$ surface concentration comes out at more than twice that in the high-pressure specimen. In the former case, there is thus an additional doping. Among the different possibly active point defects, cations in substitution appear mor likely candidates than oxygen vacancies, in qualitative agreement with previous observations\cite{Willmott2007}. 

In summary, we have thus shown that Ti ions bearing valence 3+ are indeed present at the polar interface between LAO and TiO$_2$-terminated (001) STO, but we have also shown, for the first time to our knowledge, that their concentration depends on growth conditions, so that their origin is, at least partially, extrinsic.

\acknowledgments
We thank M. Bibes (IEF-CNRS Universit\'e Paris-Sud-XI) for fruitfull discussions and M. Tenc\'e (LPS-CNRS Universit\'e Paris-Sud-XI) for his help with STEM-EELS.

\end{document}